\documentclass[conference]{IEEEtran}
\IEEEoverridecommandlockouts
\usepackage{cite}
\usepackage{amsmath,amssymb,amsfonts}
\usepackage{algorithmic}
\usepackage{graphicx}
\usepackage{textcomp}
\usepackage{xcolor}
\usepackage{bm}
\usepackage{mathptmx}
\allowdisplaybreaks[4]
\makeatletter
\newcommand{\linebreakand}{%
  \end{@IEEEauthorhalign}
  \hfill\mbox{}\par
  \mbox{}\hfill\begin{@IEEEauthorhalign}
}
\makeatother

\DeclareMathAlphabet{\pazocal}{OMS}{zplm}{m}{n}
\let\Phi\varPhi
\graphicspath{{./figures/}}
\def\BibTeX{{\rm B\kern-.05em{\sc i\kern-.025em b}\kern-.08em
    T\kern-.1667em\lower.7ex\hbox{E}\kern-.125emX}}
\begin{document}

\title{Power Control of Grid-Forming Converters Based on Full-State Feedback}

\author{\IEEEauthorblockN{Meng Chen, Dao Zhou, and Frede Blaabjerg}
\IEEEauthorblockA{\textit{AAU Energy}\\\textit{Aalborg University}\\ Aalborg, Denmark}
mche@energy.aau.dk, zda@energy.aau.dk, fbl@et.aau.dk
\thanks{This work was supported by the Reliable Power Electronic-Based Power System (REPEPS) project at the Department of Energy Technology, Aalborg University as part of the Villum Investigator Program funded by the Villum Foundation.}
}

\maketitle

\begin{abstract}
The active and reactive power controllers of grid-forming converters are traditionally designed separately, which relies on the assumption of loop decoupling. This paper proposes a full-state feedback control for the power loops of grid-forming converters. First, the power loops are modeled considering their natural coupling, which, therefore, can apply to all kinds of line impedance, i.e., resistive, inductive, or complex. Then a full-state feedback control design is used. By this way, the eigenvalues of the system can be arbitrarily placed to any positions in the timescale of power loops. Therefore, the parameters can be directly chosen by the predefined specifications. A step-by-step parameters design procedure is also given in this paper. Experimental results verify the proposed method.
\end{abstract}


\section{Introduction}
The control-based grid-forming converters are becoming vital to the power system due to their ability to help stabilize the frequency and voltage. The control system of the grid-forming converters is typically complicated with multiple loops, e.g., cascaded voltage and current loops, active and reactive power loops. To simply the analysis and design, the cascaded loops are usually designed with higher bandwidths than those of the power loops. As a result, the quick dynamics of the cascaded loops can be neglected when focusing on the power loops \cite{Rosso2021}. Moreover, the frequency and voltage are also assumed to be independently controlled by the active and reactive powers, respectively \cite{Tayyebi2020}. Popular control strategies of such grid-forming converters are the droop control and the virtual synchronous generator control \cite{Chen2021b,Zhang2021,Chen2022a}.

However, the coupling between the active and reactive power loops is highly related to the line impedance. In a resistive grid, the frequency is more influenced by the reactive power, and the voltage is more related to the active power, which is opposite with the condition of a inductive grid \cite{Deng2021}. Furthermore, for a complex impedance, the frequency, voltage, active and reactive powers have strong relations, where the active and reactive power loops are not supposed to be decoupled. The virtual impedance is a typical method to strengthen the decoupling between the active and reactive powers, which should be designed carefully to obtain a favorable performance \cite{Ahmed2022}. Nevertheless, even in a inductive grid, the decoupling between the active and reactive power loops may need additional restraints \cite{Wu2016}. Meanwhile, the decoupling is just an approximate rather than exact result. In conclusion, the decoupling-based power control may not provide a robust structure for the grid-forming converters. To get rid of the assumption of power decoupling, the multivariable feedback-based grid-forming converters have been proposed, which can simultaneously tunes both the active and reactive power loops \cite{Huang2020,Chen2022}.

The parameters design of the power control is of great importance to the grid-forming converters in order to well behave the advantages and provide a superior performance. The traditional method is based on the classic control, e.g., root locus analysis, frequency analysis, etc \cite{Chen2019,Chen2021}. These methods may be complicated when dealing with multiple loops and parameters like coupled active and reactive power loops in the grid-forming converters as they can only study the single-input single-output system with one adjustable parameters at a time.

In this paper, a full-state feedback based-power control is proposed for the grid-forming converters. With this method, the following advantages can be achieved.
\begin{enumerate}
	\item The model considers the natural coupling between the active and reactive power loops without decoupling assumptions. Therefore, the control structure is effective to any kinds of line impedance, i.e., resistive, inductive, or complex.
	\item The close-loop eigenvalues of the power loops can be arbitrarily placed to any positions in the timescale of power loops.
         \item A step-by-step parameters design procedure based on the proposed method is provided.
\end{enumerate}
 
The remainder of paper is organized as follows. The model of the grid-forming converter power loops is built in Section II. In Section III, the details of the full-state feedback-based power control is given. In Section IV, experimental results are shown, and finally, conclusions are given in Section V.

\section{Modeling of Grid-Forming Converters Power Loops}

\begin{figure}[!t]
\centering
\includegraphics[width=\columnwidth]{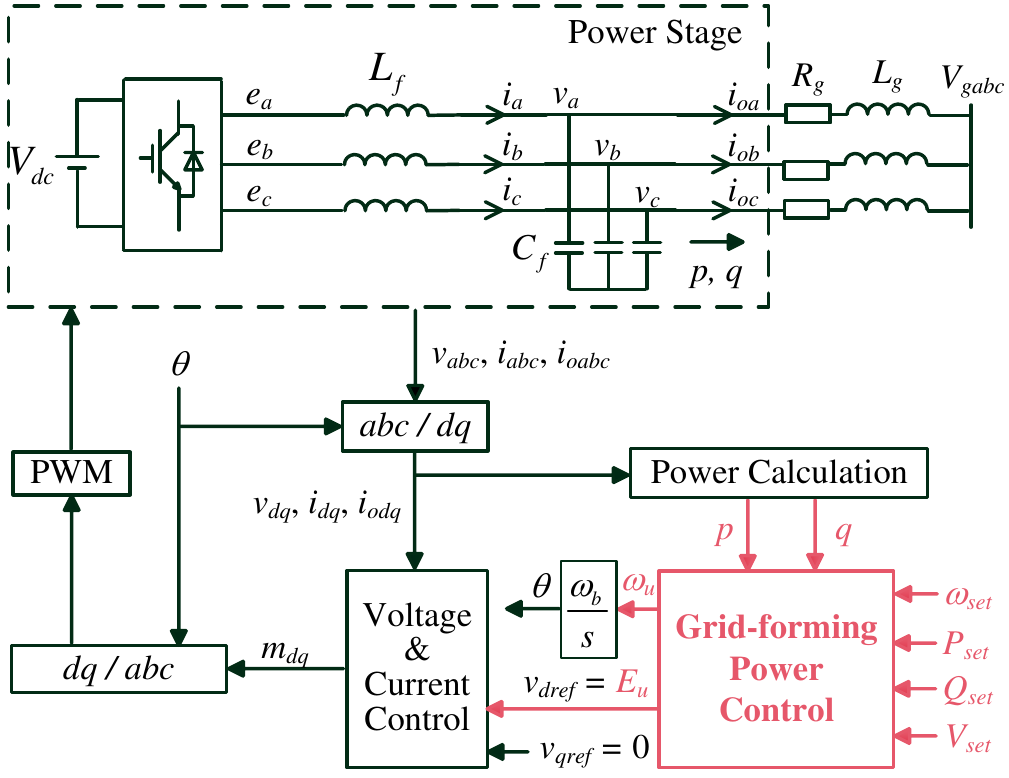}
\caption{General configuration of grid-forming converters.}
\label{VSC}
\end{figure}

Fig. \ref{VSC} shows the general configuration of a power converter controlled by the grid-forming power controller with the cascaded voltage and current loops. A three-phase converter is connected to the power grid via an LC filter, where $L_f$ and $C_f$ are the filter inductor and capacitor, respectively. $L_g$ and $R_g$ are the inductor and resistor of the line to the power grid, where both of them will be included into the modeling to represent all kinds of line impedance. The grid-forming power control copes with the output active and reactive powers of the converter, $p$ and $q$, to provide the frequency and voltage references, i.e., $\omega_u$ and $E_u$.

When considering a general line impedance, the output active and reactive powers of the power converter can be expressed as \cite{Mohammed2022}
\begin{align}
\label{p}
&p=\frac{V^2R_g+VV_g(X_gsin\delta-R_gcos\delta)}{R_g^2+X_g^2}\\
&q=\frac{V^2X_g-VV_g(R_gsin\delta+X_gcos\delta)}{R_g^2+X_g^2}
\end{align}
where $V$ and $V_g$ are the voltage magnitudes of the capacitor and grid, respectively. Moreover, $\delta$ is the power angle, which is defined as
\begin{align}
\label{delta}
\dot\delta=\omega_b\omega-\omega_b\omega_g
\end{align}
where $\omega$ and $\omega_g$ are the voltage frequencies of the capacitor and grid, and $\omega_b$ is the base value of the frequency. 

According to (\ref{p})-(\ref{delta}), the small-signal model of the output powers can be derived as
\begin{align}
\label{deltap}
&\Delta p=K_{p\delta}\Delta\delta+K_{pV}\Delta V\\
&\Delta q=K_{q\delta}\Delta\delta+K_{qV}\Delta V\\
\label{deltadelta}
&\Delta\dot\delta=\omega_b\Delta\omega
\end{align}
where
\begin{align}
\label{kpdelta}
&K_{p\delta}=\left.\frac{\partial p}{\partial\delta}\right|_{\delta_0,V_0}=\frac{V_0V_g(R_gsin\delta_0+X_gcos\delta_0)}{R_g^2+X_g^2}\\
&K_{pV}=\left.\frac{\partial p}{\partial V}\right|_{\delta_0,V_0}=\frac{2V_0R_g+V_g(X_gsin\delta_0-R_gcos\delta_0)}{R_g^2+X_g^2}\\
&K_{q\delta}=\left.\frac{\partial q}{\partial\delta}\right|_{\delta_0,V_0}=\frac{V_0V_g(X_gsin\delta_0-R_gcos\delta_0)}{R_g^2+X_g^2}\\
\label{kqv}
&K_{qV}=\left.\frac{\partial q}{\partial V}\right|_{\delta_0,V_0}=\frac{2V_0X_g-V_g(R_gsin\delta_0+X_gcos\delta_0)}{R_g^2+X_g^2}
\end{align}
and the subscript "0" represent the variables corresponding to the used steady-state operation point to linearize the model. Apparently, both the active and reactive powers are related not only to $\Delta\delta$ but also to $\Delta V$ if considering a general line impedance.

Due to the much larger bandwidths of the cascaded loops, their quick dynamics can be neglected, which yields
\begin{align}
\label{cascaded}
\left[\begin{matrix}
	\Delta\omega\\\Delta V
\end{matrix}\right]=\left[\begin{matrix}
	\Delta\omega_u\\\Delta E_u
\end{matrix}\right]
\end{align}

\begin{figure}[!t]
\centering
\includegraphics[width=\columnwidth]{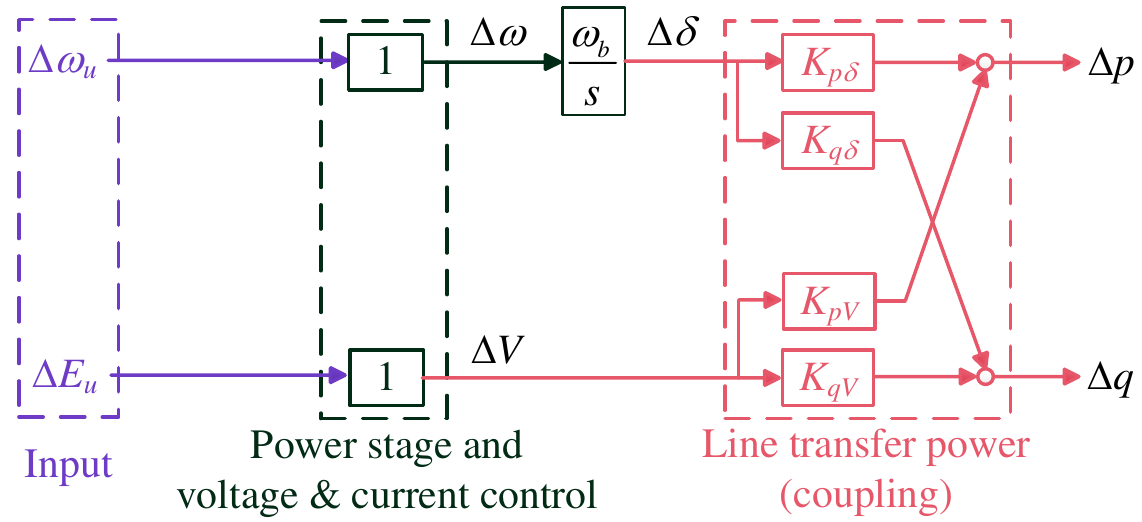}
\caption{Open-loop small-signal model of power loops.}
\label{open_loop}
\end{figure}

Combining (\ref{deltap})-(\ref{deltadelta}) and (\ref{cascaded}), the open-loop small-signal model of the grid-forming converter power loops can be graphically shown in Fig. \ref{open_loop}. As observed, the active and reactive power loops are coupled with each other for a general line impedance.

\begin{figure}[!t]
\centering
\includegraphics[width=\columnwidth]{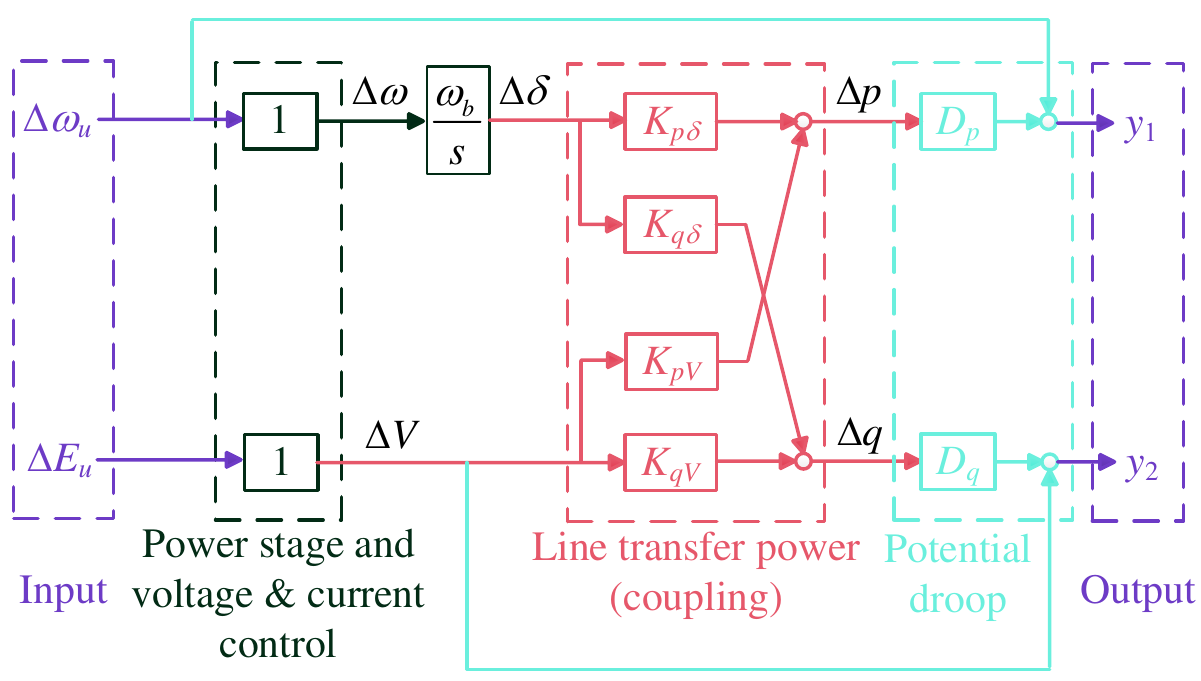}
\caption{Extended open-loop small-signal model of power loops with droop characteristics.}
\label{extended_open}
\end{figure}

Usually, a droop control is preferred to be included in the grid-forming control due to the potential islanded operation after various disturbances. A typical choice of the droop characteristics can be expressed as
\begin{align}
\label{p_droop}
&\omega_u-\omega_{set}=D_p(P_{set}-p)\\
\label{q_droop}
&V-V_{set}=D_q(Q_{set}-q)
\end{align}
where $D_p$ and $D_q$ are the droop coefficients, the subscript "set" represents the variables corresponding to the set-point. Motivated from this, two new small-signal outputs are defined as
\begin{align}
\label{y}
\left[\begin{matrix}
y_1\\y_2
\end{matrix}\right]=\left[\begin{matrix}
\Delta\omega_u+D_p\Delta p\\\Delta V+D_q\Delta q
\end{matrix}\right]
\end{align}
where their references are defined as
\begin{align}
\label{ref}
\left[\begin{matrix}
y_{1ref}\\y_{2ref}
\end{matrix}\right]=\left[\begin{matrix}
\Delta\omega_{set}+D_p\Delta P_{set}\\\Delta V_{set}+D_q\Delta Q_{set}
\end{matrix}\right]
\end{align}
By including (\ref{y}), the open-loop small-signal model in Fig. \ref{open_loop} is extended to Fig. \ref{extended_open}. As shown, it is a coupled two-input two-output system, where the state-space representation can be derived as
\begin{align}
\label{open_state}
&\Delta\dot\delta=\left[\begin{matrix}
\omega_b&0
\end{matrix}\right]\left[\begin{matrix}
\Delta\omega_u\\\Delta E_u
\end{matrix}\right]\\
\label{open_output}
&\left[\begin{matrix}
y_1\\y_2
\end{matrix}\right]=\left[\begin{matrix}
D_pk_{p\delta}\\D_qk_{q\delta}
\end{matrix}\right]\Delta\delta+\left[\begin{matrix}
1&D_pk_{pV}\\0&1+D_qk_{qV}
\end{matrix}\right]\left[\begin{matrix}
\Delta\omega_u\\\Delta E_u
\end{matrix}\right]
\end{align}

Therefore, the control target for the is to design proper close-loop controllers so that the the outputs $\left[\begin{matrix} y_1&y_2\end{matrix}\right]^T$ of the open-loop system represented by (\ref{open_state}) and (\ref{open_output}) can track the references $\left[\begin{matrix} y_{ref1}&y_{ref2}\end{matrix}\right]^T$ with zero steady-state errors by favorable dynamics. In the following, it will show that the full-state feedback can well solve the problem.

\section{Full-State Feedback Control Design}

\subsection{Control Law Construction}
To evaluate the errors between the references and the outputs, the following variables are defined.
\begin{align}
\left[\begin{matrix}
e_1\\e_2
\end{matrix}\right]=\left[\begin{matrix}
y_1-y_{1ref}\\y_2-y_{2ref}
\end{matrix}\right]
\end{align}

As the references of (\ref{ref}) are usually step disturbances, the derivatives of the errors can be derived as
\begin{align}
\label{de_definition}
\left[\begin{matrix}
\dot e_1\\\dot e_2
\end{matrix}\right]=\left[\begin{matrix}
\dot y_1\\\dot y_2
\end{matrix}\right]
\end{align}

Placing (\ref{open_output}) into (\ref{de_definition}) yields
\begin{align}
\label{de_intermediate}
\left[\begin{matrix}
\dot e_1\\\dot e_2
\end{matrix}\right]=\left[\begin{matrix}
D_pk_{p\delta}\\D_qk_{q\delta}
\end{matrix}\right]\Delta\dot\delta+\left[\begin{matrix}
1&D_pk_{pV}\\0&1+D_qk_{qV}
\end{matrix}\right]\left[\begin{matrix}
\Delta\dot\omega_u\\\Delta\dot E_u
\end{matrix}\right]
\end{align}

Define the following intermediate variables
\begin{align}
\label{z_definition}
z=\Delta\dot\delta,~\left[\begin{matrix}
u_1\\u_2
\end{matrix}\right]=\left[\begin{matrix}
\Delta\dot\omega_u\\\Delta\dot E_u
\end{matrix}\right]\end{align}
where (\ref{de_intermediate}) can, therefore, be rewritten as
\begin{align}\left[\begin{matrix}
\label{e_state}
\dot e_1\\\dot e_2
\end{matrix}\right]=\left[\begin{matrix}
D_pk_{p\delta}\\D_qk_{q\delta}
\end{matrix}\right]z+\left[\begin{matrix}
1&D_pk_{pV}\\0&1+D_qk_{qV}
\end{matrix}\right]\left[\begin{matrix}
u_1\\u_2
\end{matrix}\right]
\end{align}

Moreover, according to the definition of $z$ in (\ref{z_definition}), its derivative is
\begin{align}
\dot z=\Delta\ddot\delta
\end{align}
which can, combining with (\ref{open_state}), be rewritten as
\begin{align}
\label{dz}
\dot z=\left[\begin{matrix}
\omega_b&0
\end{matrix}\right]\left[\begin{matrix}
\Delta\dot\omega_u\\\Delta\dot E_u
\end{matrix}\right]
\end{align}

Afterwards, the dynamics of $z$ can be determined by placing (\ref{z_definition}) into (\ref{dz}) as
\begin{align}
\label{z_state}
\dot z=\left[\begin{matrix}
\omega_b&0
\end{matrix}\right]\left[\begin{matrix}
u_1\\u_2
\end{matrix}\right]
\end{align}

Therefore, the extended state differential equations of the power loops can, by taking (\ref{e_state}) and (\ref{z_state}) together, be derived  as
\begin{align}
\label{extended_open_state}
\left[\begin{matrix}
\dot e_1\\\dot e_2\\\dot z
\end{matrix}\right]=
\bm A\left[\begin{matrix}e_1\\e_2\\z\end{matrix}\right]+\bm B\left[\begin{matrix}
u_1\\u_2
\end{matrix}\right]\end{align}
where
\begin{align}
\label{A}
\bm A=\left[\begin{matrix}
0&0&D_pk_{p\delta}\\0&0&D_qk_{q\delta}\\0&0&0
\end{matrix}\right],~\bm B=\left[\begin{matrix}
1&D_pk_{pV}\\0&1+D_qk_{qV}\\\omega_b&0
\end{matrix}\right]
\end{align}

According to the full-state feedback theory \cite{Dorf2010}, the eigenvalues of the close-loop system can be placed to anywhere by the following control law if the open-loop system ($\bm A$, $\bm B$) is completely controllable.
\begin{align}
\label{control_law}
\left[\begin{matrix}
	u_1\\u_2
\end{matrix}\right]=-\bm K\left[\begin{matrix}
	e_1\\e_2\\z
\end{matrix}\right]
\end{align}
where 
\begin{align}
\label{k}
\bm K=\left[\begin{matrix}
	k_{11}&k_{12}&k_{13}\\
	k_{21}&k_{22}&k_{23}
\end{matrix}\right]
\end{align}
and the open-loop system ($\bm A$, $\bm B$) is completely controllable if the following defined controllability matrix $\bm P$ has
\begin{align}
\label{rankp}
rank(\bm P)=3
\end{align}
and
\begin{align}
\label{P}
\bm P&\triangleq\left[
\begin{matrix}
\bm B&\bm {AB}&\bm {A^2B}
\end{matrix}
\right]\notag\\
&=\left[\begin{matrix}
1&D_pk_{pV}&\omega_bD_pk_{p\delta}&0&0&0\\
0&1+D_qk_{qV}&\omega_bD_qk_{q\delta}&0&0&0\\
\omega_b&0&0&0&0&0
\end{matrix}
\right]
\end{align}

Finally, by placing (\ref{control_law}) back into (\ref{z_definition}), the actual inputs provided by the grid-forming power control can be derived as
\begin{align}
\left[\begin{matrix}
	\Delta\omega_u\\\Delta E_u
\end{matrix}\right]=\int_0^t\left(\left[\begin{matrix}
k_{11}&k_{12}\\k_{21}&k_{22}
\end{matrix}\right]\left[\begin{matrix}
-e_1\\-e_2
\end{matrix}\right]\right)d\tau-\left[\begin{matrix}
k_{13}\\k_{23}
\end{matrix}\right]\Delta\delta
\end{align}

\begin{figure*}[!t]
	\centering
	\includegraphics[width=\textwidth]{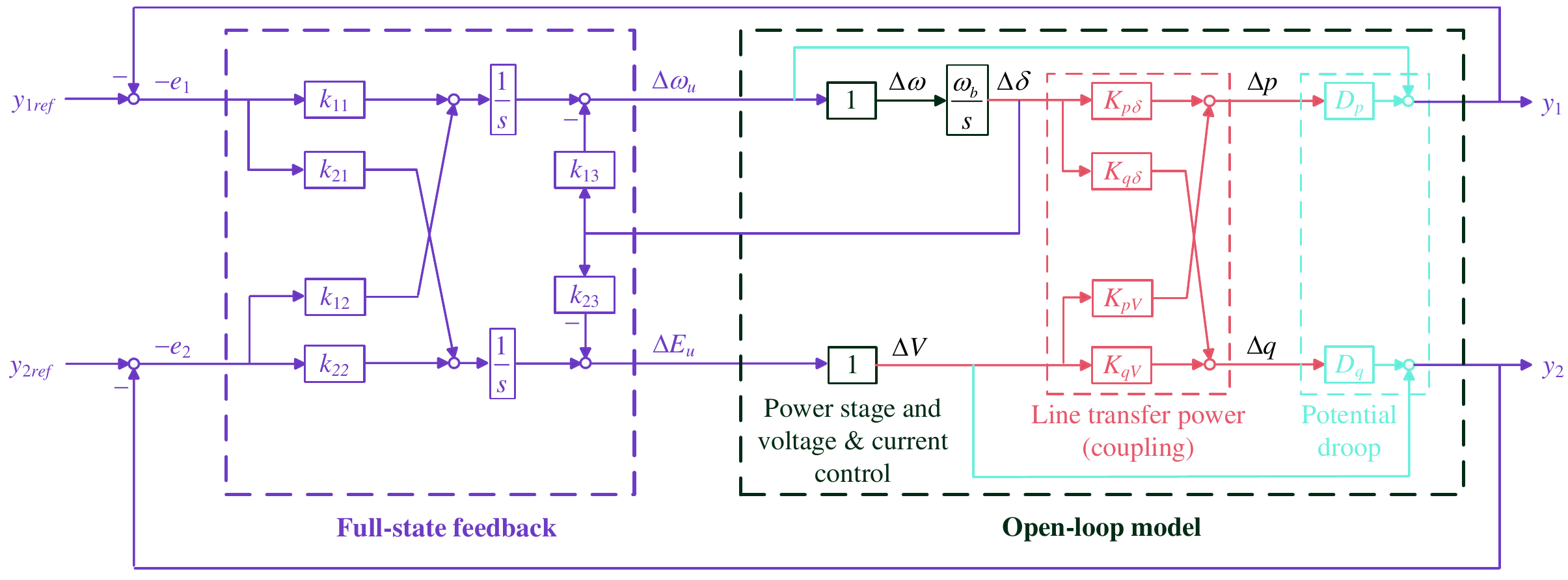}
	\caption{Close-loop small-signal model of proposed grid-forming converter.}
\label{close_loop}
\end{figure*}

Therefore, the complete close-loop small-signal model of the proposed grid-forming converter can be shown in Fig. \ref{close_loop}.

\subsection{Parameters Design}

By applying the full-state feedback control law of (\ref{control_law}), the close-loop dynamics of (\ref{extended_open_state}) is derived as
\begin{align}
\left[\begin{matrix}
\dot e_1\\\dot e_2\\\dot z
\end{matrix}\right]=
(\bm A-\bm{BK})\left[\begin{matrix}e_1\\e_2\\z\end{matrix}\right]
\end{align}
where its characteristic equation is
\begin{align}
\label{char}
	\left|\lambda\bm{I}-\bm A+\bm{BK}\right|=0
\end{align}

The characteristic equation (\ref{char}) has three eigenvalues, which can be placed in anywhere. As a reasonable choice, we choose a pair of complex eigenvalue as the dominant ones and a real eigenvalue, which is far away from the dominant eigenvalues. Therefore, the characteristic equation should has the following form
\begin{align}
	(\lambda+a)(\lambda^2+2\xi\omega_n\lambda+\omega_n^2)=0
\end{align}
where $-a$ is the chosen real eigenvalue, $\xi$ and $\omega_n$ are the damping ratio and natural frequency of the chosen complex eigenvalues. Thereafter, the parameters (\ref{k}) can be solved by
\begin{align}
\label{solve_k}
	\left|\lambda\bm{I}-\bm A+\bm{BK}\right|\equiv(\lambda+a)(\lambda^2+2\xi\omega_n\lambda+\omega_n^2)
\end{align}

It should be mentioned that (\ref{rankp}) guarantees that (\ref{solve_k}) must have solutions. Meanwhile, $\xi$ and $\omega_n$ are directly related to the time domain performance, e.g., overshoot, settle time, etc., which can, therefore, be used to determine the dominant complex eigenvalues.

According to the aforementioned discussion, a step-by-step parameter design procedure can be given as follows.
\begin{itemize}
	\item step 1: Choose the steady-state operation point to linearize the system.
	\item step 2: Calculate $K_{p\delta}$, $K_{pV}$, $K_{q\delta}$, and $K_{qV}$ of the chosen steady-state operation point based on (\ref{kpdelta})-(\ref{kqv}).
	\item step 3: Calculate the matrices $\bm A$ and $\bm B$ of (\ref{A}).
	\item step 4: Check the controllability according to (\ref{rankp}) and (\ref{P}).
	\item step 5: Choose proper eigenvalues according to the requirements of the time domain performance.
	\item step 6: Solve (\ref{solve_k}) to obtain the parameter matrix $\bm K$.
\end{itemize}    

\section{Experimental Validation}

\begin{figure}[!t]
\centering
\includegraphics[width=\columnwidth]{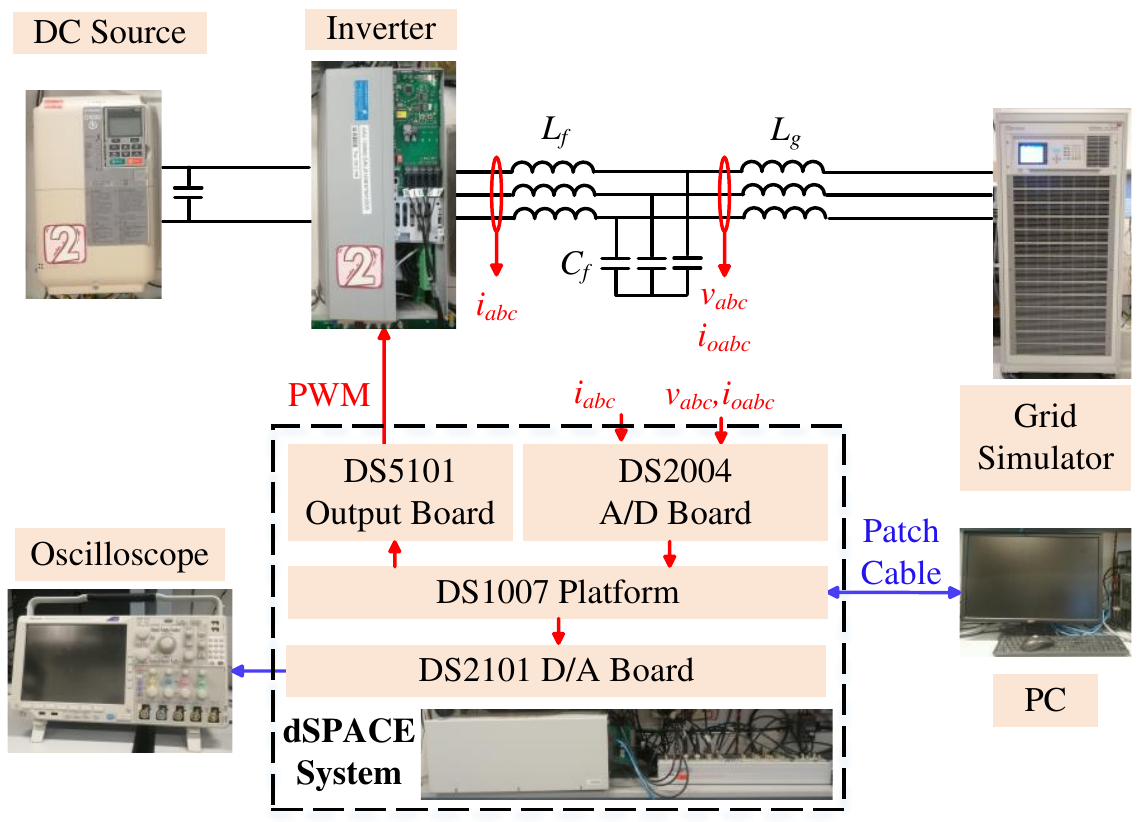}
\caption{Experimental configuration of grid-forming converter.}
\label{setup}
\vspace{-10pt}
\end{figure}

To verify the proposed full-state feedback control for power loops of the grid-forming converter as well as the step-by-step parameter design procedure, this section will present corresponding experimental results. The configuration of the setup is shown in Fig. \ref{setup}, and the key parameters are given in Table \ref{parameter}.

\begin{table}[!t]
	\renewcommand{\arraystretch}{1.3}
	\caption{Parameters of Experimental Setups}
	\centering
	\label{parameter}
	\resizebox{\columnwidth}{!}{
		\begin{tabular}{c l c}
			\hline\hline \\[-3mm]
			Symbol & Description & Value  \\ \hline
			$f_n$  & Nominal frequency & $100\pi$ rad/s \\
			$S_n $ & Nominal power &  5 kW  \\ 
			$V_n $ & Nominal line-to-line RMS voltage & 380 V \\
			$f_{sw}$ & Switching frequency & 10 kHz \\
			$\omega_g $ & Grid frequency & $100\pi$ rad/s (1 p.u.)  \\
			$V_g $ & Line-to-line RMS grid voltage & 380 V (1 p.u.) \\
			$L_g$  & Line inductor & 8 mH (0.087 p.u.)\\
			$C_f$ & Filter capacitor & 5 $\mu$F (0.0454 p.u.)\\   
			$L_f$ & Filter inductor & 3 mH (0.0326 p.u.)\\
			$D_p$ & Droop coefficient of $P$-$f$ regulation & 0.01 p.u. \\ 
			$D_q$ &Droop coefficient of $Q$-$V$ regulation & 0.05 p.u.\\
			$\omega_{set}$&Frequency reference & 1 p.u. \\
			$P_{set}$ & Active power reference & 0.5 p.u.\\
			$Q_{set}$ & Reactive power reference & 0 p.u.\\
			$V_{set}$ & Voltage magnitude reference & 1 p.u.\\[1.4ex]
			\hline\hline
		\end{tabular}
	}
\end{table}

The step-by-step parameter design procedure is applied as follows.
\begin{itemize}
	\item step 1: Choose the steady-state operation point to linearize the system.

		The steady-state operation point with the parameters in Table \ref{parameter} is ($\delta_0$, $V_0$) = (0.0435, 0.9997).
	\item step 2: Calculate $K_{p\delta}$, $K_{pV}$, $K_{q\delta}$, and $K_{qV}$ of the chosen steady-state operation point based on (\ref{kpdelta})-(\ref{kqv}).

		The corresponding parameters are calculated as ($K_{p\delta}$, $K_{pV}$, $K_{q\delta}$, $K_{qV}$) = (11.4761, 0.5002, 0.5, 11,4939).
	\item step 3: Calculate the matrices $\bm A$ and $\bm B$ of (\ref{A}).

		The corresponding matrices are calculated as
		\begin{align}
		\bm A=\left[\begin{matrix}
			0&0&0.1148\\
			0&0&0.025\\
			0&0&0
		\end{matrix}\right],~\bm B=\left[\begin{matrix}
			1&0.005\\
			0&1.5747\\
			314.1593&0
		\end{matrix}\right]
		\end{align}
	\item step 4: Check the controllability according to (\ref{rankp}) and (\ref{P}).

		The controllability matrix is calculated as
		\begin{align}
		\bm P=\left[\begin{matrix}
			1&0.005&36.0533&0&0&0\\
			0&1.5747&7.854&0&0&0\\
			314.1593&0&0&0&0&0
		\end{matrix}\right]
		\end{align}
		where it is easy to check that $rank(\bm P)=3$. Therefore, the system ($\bm A$, $\bm B$) is completely controllable.
	\item step 5: Choose proper eigenvalues according to the requirements of the time domain performance.

As an example, in this paper, we consider the percent overshoot ($P.O.$) and settle time ($T_s$) as the performance indices, where they are related to $\xi$ and $\omega_n$ by the following equations \cite{Ogata2009}.
\begin{align}
&P.O.=e^{-(\xi/\sqrt{1-\xi^2})}\times100\%\\
&T_s=\frac{4}{\xi\omega_n}
\end{align}
For comparison, this paper choose four cases as shown in Table \ref{case}, where the third eigenvalue $a$ is always placed far away from the dominant complex eigenvalues. The corresponding positions of the dominant eigenvalues are shown in Fig. \ref{eig}.

\begin{table}[!t]
	\renewcommand{\arraystretch}{1.3}
	\caption{Studied Cases to Place Eigenvalues}
	\centering
	\label{case}
	\resizebox{\columnwidth}{!}{
		\begin{tabular}{c l cc}
			\hline\hline \\[-3mm]
			Cases & Damping Ratio $\xi$ & Settle Time $T_s$  & Third eigenvalue $a$\\ \hline
			1  & 0.4 & 1 s & -20\\
			2 & 0.4 &  2 s & -20  \\ 
			3 & 0.707 & 1 s &-20 \\
			4 & 0.707 & 2 s & -20 \\[1.4ex]
			\hline\hline
		\end{tabular}
	}
\end{table}

\begin{figure}[!t]
	\centering
	\includegraphics[width=\columnwidth]{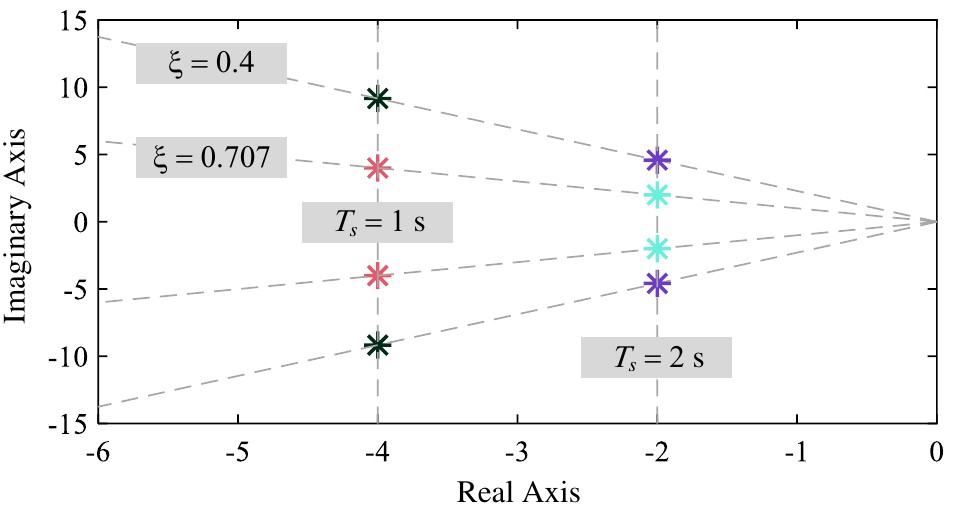}
	\caption{Chosen dominant complex eigenvalues.}
\label{eig}
\end{figure}

	\item step 6: Solve (\ref{solve_k}) to obtain the parameter matrix $\bm K$.

	The designed parameters corresponding to the four cases are listed in Table \ref{designed_parameter}.
\end{itemize}   

\begin{table}[!t]
	\renewcommand{\arraystretch}{1.3}
	\caption{Designed parameters of the corresponding cases}
	\centering
	\label{designed_parameter}
	\resizebox{0.9\columnwidth}{!}{
		\begin{tabular}{c c c c c}
			\hline\hline \\[-3mm]
			Parameters & Case 1 & Case 2 & Case 3 & Case 4  \\ \hline
			$k_{11}$  & 2.7756 & 0.6939 &0.8885&0.2221\\
			$k_{12}$ & -0.0088 &  -0.0022 & -0.0028&-0.0007\\ 
			$k_{13}$ & 0.0166 & 0.0105 &0.0226&0.012\\
			$k_{21}$ & 0.0367 & 0.0389 &0.0385&0.0399\\
			$k_{22}$ & 12.7007 & 12.7007& 12.7007&12.7007\\
			$k_{23}$ & 0.0161 & 0.0161 &0.0161&0.0161\\[1.4ex]
			\hline\hline
		\end{tabular}
	}
\end{table}

\begin{figure*}[!t]
	\centering
	\includegraphics[width=0.88\textwidth]{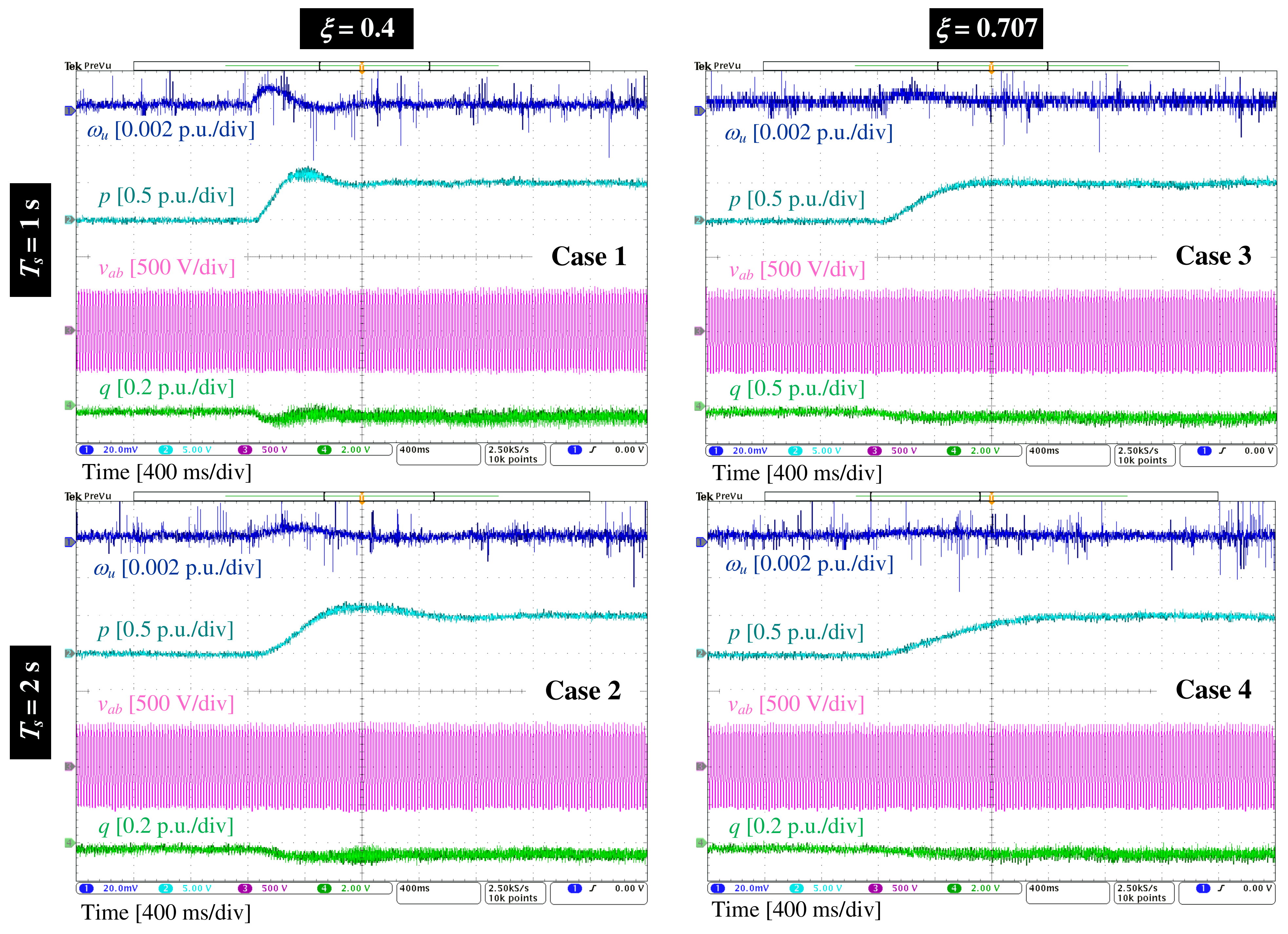}
	\caption{Experimental results of studies cased when $P_{ref}$ steps from 0.5 p.u. to 1 p.u.}
	\label{experiment}
\end{figure*}

Fig. \ref{experiment} presents the experimental comparisons of the studied cases when $P{set}$ steps from 0.5 p.u. to 1 p.u.. As shown, when choosing a large damping ratio ($\xi=0.707$ in Case 3 and Case 4), the dynamics have smaller $P.O.$ than those with a small damping ratio ($\xi=0.4$ in Case 1 and Case 2). Meanwhile, when choosing a small settle time ($T_s=1$s in Case 1 and Case 3), the systems can reach to the steady-state quicker than those with a large settle time ($T_s=2$s in Case 2 and Case 4). Fig. \ref{experiment} proves that the proposed full-state back control structure and parameter design method are effective to regulate the power loops of the grid-forming converter to satisfy the predefined time domain performance.

\section{Conclusion}
This paper proposes a new control structure of the grid-forming converter power loops based on the full-state feedback control. The modeling considers a general line impedance and the coupling between the power loops, which makes that the proposed control can be applied to a inductive, resistive, or complex grid line. A step-by-step parameter design procedure is given as well, where the eigenvalues can be placed anywhere within the timescale of the power loops. By this way, the control parameters can be directly calculated based on the predefined time domain indices, which are verified by the experimental results.

\bibliographystyle{IEEEtran}
\bibliography{IEEEabrv,ECCEAsia}
\end{document}